# Ultrafast Successive Control and Electrical Detection of Long-Lived Valley Polarization in Room-Temperature Germanium

Ami M. Shirai[1]*, Yuta Murotani[1‡], Shunta Aoki[1], Shunsuke A. Sato[2,3], Tomohiro Fujimoto[1§], Kota Aikyo[1], Changsu Kim[1], Hidefumi Akiyama[1], Jun Yoshinobu[1], Toshiya Ideue[1], and Ryusuke Matsunaga[1†]

[1]*The Institute for Solid State Physics, The University of Tokyo, Kashiwa, Chiba 277-8581, Japan*

[2]*Department of Physics, Tohoku University, Sendai 980-8578, Japan*

[3]*Max Planck Institute for the Structure and Dynamics of Matter, Luruper Chaussee 149, 22761 Hamburg, Germany*

*shirai-ami@issp.u-tokyo.ac.jp

†matsunaga@issp.u-tokyo.ac.jp

‡Present address: Institute for Chemical Research, Kyoto University, Uji, Kyoto 611-0011, Japan

§Present address: NTT Basic Research Laboratories, NTT Corporation, Atsugi, Kanagawa 243-0198, Japan

## Abstract

Germanium (Ge) hosts four degenerate valleys in the conduction band minima, and their highly anisotropic effective masses offer a valley degree of freedom in transport phenomena via anisotropic conductivity. Using time- and polarization-resolved terahertz spectroscopy combined with deformation potential theory, we reveal a valley relaxation time reaching 10 ps near the band edge of Ge(001) at room temperature, long enough to detect the valley polarization in electrical transport measurements. We also demonstrate sub-picosecond complete reversal of valley polarization by off-resonant light pulses at extremely low energy consumption. These results pave the way for room-temperature valleytronics based on bulk indirect-gap semiconductors compatible with existing electronics.

Controlling the internal degrees of freedom of carriers is crucial for developing next-generation, low-power, high-efficiency electronic devices. While spintronics uses the spin degree of freedom, valleytronics exploits the energy band minima in momentum space, or the valley degree of freedom, which has recently attracted considerable attention. A prime example is monolayer transition metal dichalcogenides (TMDCs), represented by $MoS_2$ [1,2]. These materials exhibit degenerate bandgaps in the K and K' valleys, allowing valley-selective carrier excitation using circularly polarized light via optical selection rules for interband transitions [3–5]. Furthermore, a long valley polarization lifetime was initially expected because intervalley scattering requires a large momentum transfer with a simultaneous spin flip due to the spin-valley locking [6]. Recently, off-resonant [7–9] and coherent control [10] of valley polarization have also been reported. To realize practical electronic device applications in valleytronics, the detection of valley polarization through transport in electric circuits is indispensable, which is enabled by the valley Hall effect [11] owing to the valley-dependent Berry curvatures [6].

However, TMDC-based valleytronics faces critical challenges in valley lifetime, conversion into electric current, and compatibility with existing semiconductor electronics. At room temperature, valley polarization in $MoS_2$ is lost within only 75 fs [12], owing to the strong interaction with zone-edge phonons [13], short-range interaction with defects and impurities, and exchange interaction between an electron and a hole [14–16]. Furthermore, direct observation of the valley Hall effect is notoriously difficult, so that very few studies have succeeded in directly demonstrating this effect since the initial pioneering reports [11,17]. Moreover, obtaining high-quality samples requires encapsulation in hexagonal boron nitride (h-BN) [18], creating numerous hurdles for practical electronic device integration.

Historically, the relationship between the valley structure and transport responses has been investigated in conventional indirect-bandgap semiconductors, such as silicon (Si) and germanium (Ge) [19]. Figure 1(a) illustrates an example of a fourfold degenerate valley structure at the L point, forming conduction band (CB) minima in Ge. In these materials, because of the phonon-assisted indirect gap transition and highly anisotropic energy dispersion of each valley, the interband transitions induced by linearly polarized light are accompanied by different excitation probabilities for each valley, thereby creating valley polarization, as shown in Fig. 1(b) [20,21]. Once valley polarization is generated, the difference between the longitudinal and transverse effective masses, $m_l^e$ and $m_t^e$, respectively, in each valley leads to anisotropic conductivity. Recently, light-induced valley polarization dynamics in Si has been investigated using terahertz (THz) pulses [22], where anisotropic conductivity can be probed as linear birefringence and dichroism in the THz frequency range. After injecting valley-polarized carriers into the band edge using near-infrared (NIR) pulses, the relaxation time of valley polarization exceeding 1.5 ps at room temperature—20 times longer than that of $MoS_2$—was revealed in the time-resolved measurements [22]. Additionally, off-resonant manipulation of valley polarization using light with a photon energy lower than the bandgap has recently been demonstrated in diamond and Si [23], highlighting its potential for applications using conventional semiconductors. Compared with the effective mass anisotropy $m_l^e/m_t^e = 4.8$ in Si, Ge exhibits an even higher anisotropy ($m_l^e/m_t^e = 19.5$) as well as a higher carrier mobility. Although the ultrafast Γ–L intervalley scattering in Ge has been intensively studied due to its strong relevance to high-performance transistors and optoelectronic devices [24–26], the valley polarization dynamics at the band bottom determined by the L–L intervalley scattering has evaded experimental observation at room temperature. A pioneering acoustoelectric experiment suggested a long valley relaxation time of approximately 100 ps in Ge at temperatures below 100 K [27]. Given its seamless integration with current semiconductor technologies, Ge is a promising candidate for future applications in valleytronic devices.

In this Letter, we explore possible valleytronics in bulk Ge at room temperature. We investigate the dynamics of valley polarization by exciting carriers at the indirect bandgap edge of bulk Ge(001) using wavelength-tunable NIR pulses and measuring the resulting anisotropic conductivity through the polarization rotation of the THz pulses [Fig. 1(c)]. Furthermore, by comparing our results with theoretical calculations incorporating deformation potentials, we successfully discriminate between the momentum relaxation dynamics of holes asymmetrically distributed near the Γ point and the valley relaxation dynamics of electrons, revealing that electrons in Ge exhibit an exceptionally long valley relaxation time of nearly 10 ps. This long lifetime enables detection via time-integrated direct current (DC) transport measurements. In addition, by accelerating electrons using a linearly polarized mid-infrared light field, we demonstrate off-resonant generation and successive, sub-picosecond switching of valley polarization with ultralow energy consumption. This study paves the way for ultrafast THz information processing based on valleytronics in indirect-gap semiconductors.

The sample was a 100-μm-thick bulk Ge crystal. All experiments were performed at room temperature. First, we conducted time-resolved THz polarimetry by employing three optical pulses: NIR pump, THz probe, and NIR gate pulses, with the pulse sequence shown in Fig. 2(a). NIR pump pulses were generated using an optical parametric amplifier with wavelengths tuned between 1650 and 1850 nm, corresponding to photon energies near the indirect bandgap but below the direct one [28]. The THz probe pulses were generated via optical rectification in a GaP crystal and detected by electro-optic sampling using 1030-nm gate pulses with a pulse duration shorter than 100 fs. The pump pulses were linearly polarized along the [110] or [$\bar{1}$10] crystal axis so that the valley polarization can be generated on the L point [Figs. 1(a) and 1(b)]. To detect the valley polarization via THz polarization rotation, the incident THz pulse was polarized along the [100] direction, which we chose as the horizontal ($x$) axis in the laboratory frame. Two wire-grid polarizers were positioned after the sample to separately detect the respective $x$- and $y$-polarization components of the transmitted THz pulse, $E_x$ and $E_y$. From the polarization rotation $E_y/E_x$, the anisotropic conductivity $\sigma_{\text{ani}} \equiv \sigma_{[110]} - \sigma_{[\bar{1}10]}$, which represents the conductivity difference between the [110] and $[\bar{1}10]$ crystal axes, was extracted. To eliminate the offsets independent of the pump polarization direction, the difference between the two $E_y$ signals for the mutually orthogonal pump polarization directions along the [110] and $[\bar{1}10]$ crystal axes was obtained (for details, see Supplemental Material [29]). Notably, not only are electrons valley polarized, but the holes are also asymmetrically distributed around the Γ point [35,36] and contribute to the anisotropic conductivity.

Figure 2(b) shows the two-dimensional (2D) plots of the change in transmitted THz field induced by the 1700 nm NIR pump as functions of the delay times of the THz probe and NIR pump pulses, $t_{\text{probe}}$ and $t_{\text{pump}}$, respectively; the left and right panels show parallel $[\Delta E_x(t_{\text{probe}}, t_{\text{pump}})]$ and perpendicular $[\Delta E_y(t_{\text{probe}}, t_{\text{pump}})]$ components to the incident field. $\Delta E_x(t_{\text{probe}}, t_{\text{pump}})$ represents the decrease in transmission due to photocarrier absorption, whereas $\Delta E_y(t_{\text{probe}}, t_{\text{pump}})$ captures the anisotropic conductivity due to the pump-induced valley polarization. In contrast to previous observations in Si [22], the diagonally oscillating signal around $t_{\text{pump}} = 0$, indicating the field-induced photogalvanic effect [37,38], is small in Ge; thus, we neglect it in the following discussion.

To analyze the anisotropic conductivity dynamics more quantitatively, $\Delta E_y$ at a fixed probe delay was measured as a function of $t_{\text{pump}}$ at various pump wavelengths. Figure 2(c) presents $\Delta E_y(t_{\text{pump}})$ at a fixed $t_{\text{probe}}$ for the 1700 nm pump at a fluence of 11 μJ/cm$^2$. Two distinct relaxation components, fast and slow, were isolated using a double-exponential fit convolved with a Gaussian instrument response function (see Supplemental Material [29]). The fast and slow components are plotted separately in Fig. 2(c). The extracted relaxation times are plotted in Fig. 2(d) against $\hbar\omega - E_g$, the excess energy above the bandgap ($E_g = 0.66$ eV). Both the fast and slow relaxation times increased with decreasing excess energy, which is accounted for by fewer scattering channels for small excess energies. Near the band edge, the fast relaxation time was 1.4 ps, whereas the slow relaxation time reached approximately 10 ps.

For comparison, we calculated the inverse of the L–L intervalley scattering rate of electrons $\tau_e^{-1}(\varepsilon)$ and the hole-phonon scattering rate $\tau_h^{-1}(\varepsilon)$ based on the deformation potential theory [34]:

$$\tau_e^{-1}(\varepsilon) = \sum_{j=\text{LA(LO),TA}} \sum_{i=\pm 1} \frac{Z\left(D_{\text{LL},j}^{e}\right)^2 (m_d^e)^{3/2}}{\sqrt{2}\pi\hbar^3\rho\omega_j^{\text{X}}} \left(N_{\text{ph}}\left(\omega_j^{\text{X}}\right) + \frac{1-i}{2}\right) \sqrt{\varepsilon + i\hbar\omega_j^{\text{X}}}, \tag{1}$$

$$\tau_h^{-1}(\varepsilon) = P_{\Gamma,\text{O}}^{h}(\varepsilon) + P_{\Gamma,\text{A}}^{h}(\varepsilon), \tag{2}$$

$$P_{\Gamma,\mathrm{O}}^{h}(\varepsilon) = \sum_{i=\pm 1} \frac{3\left(D_O^h\right)^2 \left(m_d^h\right)^{3/2}}{2^{3/2}\pi\hbar^3\rho\omega_O^\Gamma}\left(N_{\mathrm{ph}}\left(\omega_O^\Gamma\right) + \frac{1-i}{2}\right)\sqrt{\varepsilon + i\hbar\omega_O^\Gamma}\,, \tag{3}$$

$$P_{\Gamma,\mathrm{A}}^{h}(\varepsilon) = \frac{\sqrt{2}\left(D_A^h\right)^2 k_B T \left(m_d^h\right)^{3/2}}{\pi\hbar^4\rho u_l^2}\sqrt{\varepsilon}, \tag{4}$$

where $\rho$ is the mass density of Ge and $i = \pm 1$ represent phonon absorption and emission, respectively. For electrons, $Z = 3$ is the number of final valleys available for scattering; $D_{\mathrm{LL},j}^{e}$ is the L–L intervalley deformation potential for a phonon mode $j$, where the relevant phonons are LA, LO, and TA, with LA and LO phonons being degenerate; $\omega_j^{\mathrm{X}}$ is the phonon angular frequency at the X point; and $m_d^e$ is the electron density-of-states effective mass. For holes, $D_O^h$ and $D_A^h$ denote the optical- and acoustic-phonon deformation potentials near the Γ point, respectively. $\omega_O^\Gamma$ is the optical phonon angular frequency at the Γ point, $m_d^h$ is the hole density-of-states effective mass, $u_l$ is the longitudinal sound velocity, and $N_{\mathrm{ph}}(\omega)$ is the phonon occupation number, $1/(e^{\hbar\omega/k_B T} - 1)$. Because acoustic phonons near the Γ point have sufficiently small energy, emission and absorption processes were integrated into a single expression. These formulations describe the carrier-phonon coupling strength resulting from local lattice deformations. All parameters were obtained from the literature [34] (for details, see Supplemental Material [29]). The carrier-carrier and impurity scattering mechanisms were neglected, as their rates were substantially smaller than the carrier-phonon scattering rates at photogenerated carrier densities below $10^{17}/\mathrm{cm}^3$. The calculated results are presented in Fig. 2(e) as a function of the carrier kinetic energy ε for comparison with the observed relaxation times in Fig. 2(d). The kinks in $\tau_h(\varepsilon)$ near ε ≈37 meV and in $\tau_e(\varepsilon)$ near ε ≈28 meV arise from the onset of emission of zone-center optical phonons and LA phonons at the X point, respectively. The fast relaxation times in Fig. 2(d) align with $\tau_h(\varepsilon)$, corresponding to the relaxation of anisotropic conductivity arising from the asymmetric hole distribution. Accordingly, the slow relaxation component is attributed to $\tau_e(\varepsilon)$, the valley polarization lifetime of electrons. Note that the observed slow relaxation times in Fig. 2(d) are systematically longer than the calculated $\tau_e(\varepsilon)$. Especially at large excess energies, the slow relaxation times saturate whereas $\tau_e(\varepsilon)$ continues to decrease. This is reasonable because the photoexcited hot carriers would immediately start to thermalize within each valley, so that intervalley scattering effectively takes place at kinetic energies lower than the initial excess energy. Consequently, these distinct relaxation timescales allow us to distinguish the electron and hole contributions to the anisotropic conductivity. The long relaxation time of ~10 ps near the band edge is attributed to the valley-polarization relaxation of electrons, which is consistent with an extrapolation of previous low-temperature acoustoelectric experiments below 100 K to room temperature [27]. Remarkably, the valley polarization lifetime in Ge is an order of magnitude longer than that of Si [22], which can be attributed to its smaller density-of-states mass and higher mass density compared with Si (see Supplemental Material [29]).

Assuming a parabolic band dispersion, the relation between the anisotropic conductivity $\sigma_{\mathrm{ani}}$ and valley polarization ratio $P_v$ can be expressed through the Drude model as

$$\sigma_{\mathrm{ani}} = \left(\frac{1}{m_{\parallel}^{e}} - \frac{1}{m_{\perp}^{e}}\right)\frac{N_c^e e^2 \tau_D}{1 - i\omega\tau_D} P_v, \tag{5}$$

where $m_{\parallel}^{e}$ and $m_{\perp}^{e}$ are the effective masses parallel and perpendicular to the NIR polarization direction, respectively. $N_c^e$ is the electron density, $e$ is the elementary charge, and $\tau_D$ is the momentum relaxation time. By modeling wave propagation through an optically anisotropic medium via anisotropic conductivity, the relationship between the THz polarization rotation angle and valley polarization ratio was established for each photon energy (see Supplemental Material [29]). The estimated valley polarization ratio is presented in Fig. 2(f) as a function of the excess energy $\hbar\omega - E_g$. Intriguingly, the valley polarization ratio increased with the excess energy and reached 4% with an abrupt transition near the LA phonon energy of 28 meV. Because LA phonons play a dominant role in indirect transitions in Ge [39], the interband transition with the emission of an LA phonon may play a key role in inducing valley polarization.

To explore the detectability of the valley polarization in DC transport, we define a conductivity anisotropy angle $\theta_{\mathrm{ani}} \equiv \sigma_{yx}/\sigma_{xx}$ (Supplemental Material [29]), which scales with $P_v$. In analogy to the spin Hall angle $\theta_{\mathrm{H}} = \sigma_{\mathrm{H}}/\sigma_{xx}$, $\theta_{\mathrm{ani}}$ characterizes the detection efficiency of valley information through transport. We evaluated $\theta_{\mathrm{ani}}$ ≈0.013 for $P_v$ =4% (Supplemental Material [29]). Compared with $\theta_{\mathrm{H}}$ in semiconductors used for spintronics at room temperature [40] (~$10^{-4}$ for p-Si [41] and ~$10^{-3}$ for n-Ge [42]), our results show a remarkably large conductivity anisotropy angle, suggesting that the valley polarization can be detected as a time-integrated transport response. Thus, as depicted in Fig. 3(a), we replaced the THz field with a static DC bias voltage along the positive $x$ direction, and measured the resulting transverse photovoltage along the $y$ direction in a Hall bar configuration, with a lock-in amplifier referenced to the repetition rate of the pump pulses. The signal is defined as positive when the current flows along the positive $y$ direction. Linearly polarized NIR pump pulses at 1800 nm were focused between the electrodes without illuminating the contacts. The polarization angle of the NIR beam was rotated counterclockwise, as viewed along the incident beam, using a half-wave plate. Figure 3(b) shows the transverse voltage after subtracting the background offset, which reveals a clear dependence on the pump polarization angle and is well fitted by a 180° periodic sine curve. The oscillation phase and amplitude of the transverse voltage signal are plotted against the pump fluence in Figs. 3(c) and 3(d), respectively. The transverse voltage reached its maximum at a polarization angle of 45°, which was consistent with the valley-related anisotropic conductivity geometry induced by a linearly polarized NIR pump. The amplitude exhibited a nearly linear scaling at low fluences, ruling out nonlinear optical artifacts. While the transport signal started to deviate from linear scaling at higher fluences, the THz polarization rotation signal overlaid in Fig. 3(d), defined as the peak value of $\Delta E_y(t_{\mathrm{pump}})$ as in Fig. 2(c), remained linear up to higher fluences. This difference can be attributed to the distinct heating effects of the DC and THz pulse measurements. These results demonstrate that anisotropic conductivity is easily detectable as a DC transport signal owing to the long lifetime of the valley polarization in Ge, confirming the efficient conversion from the valley degree of freedom to charge transport that constitutes an indispensable building block for constructing valleytronic devices.

Furthermore, the strong dependence of the valley polarization lifetime on the carrier kinetic energy directly observed in Fig. 2(d) has another important implication for the controllability of valley polarization using an off-resonant light pulse [23]. According to the acceleration theorem, an off-resonant field oscillates the carrier momentum, imparting kinetic (ponderomotive) energy $U_p$ on the Bloch electrons. Because of the effective mass anisotropy of each valley, the effective mass along the field polarization direction differs among the valleys. This implies that the imparted kinetic energy is valley-dependent; electrons in valleys with a smaller effective mass along the field polarization direction acquire a larger kinetic energy than those in valleys with a larger effective mass, as illustrated in Fig. 4(a). This kinetic energy imbalance causes a difference in the intervalley scattering probabilities, as shown in Fig. 2(d), thereby generating valley polarization depending on the field polarization direction.

To demonstrate this off-resonant control of valley polarization in Ge, a 4-μm mid-infrared (MIR) drive pulse was applied to the sample at a delay of approximately 100 ps after the NIR pump, as represented by the pulse sequence in Fig. 4(b). The resulting valley polarization was probed using THz birefringence and dichroism. The ~100 ps delay from the NIR pump ensures that the photoexcited carriers are equally distributed among the L valleys before the arrival of the MIR pulse because it is much longer than the valley polarization lifetime (~10 ps) but much shorter than the carrier recombination time in Ge on the nanosecond scale. Figure 4(c) shows the THz polarization rotation signal $\Delta E_y$ at a fixed probe delay as a function of $t_{\mathrm{drive}}$ a fluence of 53 μJ/cm$^2$. An anisotropic conductivity signal $\Delta E_y$ induced by the MIR drive pulse was successfully observed, demonstrating the controllability of the valley polarization by off-resonant light without changing the carrier density (Supplemental Material [29]). The resulting $P_v$ was ~0.6%, corresponding to $\theta_{\mathrm{ani}}\sim 2\times 10^{-3}$, which still exceeds $\theta_H$ of n-Ge [42] and is thus also large enough for electrical detection. Furthermore, Fig. 4(d) shows the result obtained by applying dual MIR drive pulses with mutually orthogonal polarizations and a 0.87-ps interval. After the first MIR pulse generated valley polarization, the second MIR pulse clearly switched this imbalance to the other pair of valleys. The interval between the positive and negative peaks for this reversal was approximately 0.8 ps and matched the pulse delay. This sub-picosecond complete reversal highlights the potential for ultrafast and arbitrary valley manipulation in Ge, as discussed later.

To confirm the underlying mechanism of MIR-induced valley polarization, we conducted Monte Carlo simulations of electron scattering [23]. The model incorporated two processes: the momentum shift of carriers driven by the 4-μm MIR field, and the electron-phonon scattering. The latter included L–L intervalley scattering mediated by LA, LO, and TA phonons, as well as intravalley scattering mediated by optical and acoustic phonons. At each time step, we stochastically calculated the change in the electron distribution considering the inter- and intravalley scattering channels. In the intervalley process, an electron exchanged energy with a mediating optical or acoustic phonon and was randomly moved to one of the other three L valleys. In the intravalley process, an electron was scattered into a random momentum state within the same valley, and we neglected the energy of acoustic phonons. The initial carrier distribution followed the Fermi–Dirac distribution, with the chemical potential determined from the experimentally estimated carrier density. Each electron iteratively underwent the field-acceleration and scattering-determination processes with sufficiently small time steps. The MIR fluence was adjusted within the range used in the experiments. Further details are provided in the Supplemental Material [29]. Figure 4(e) shows the experimentally obtained anisotropic conductivity dynamics (upper panel) and the simulated valley polarization ratio of electrons (lower panel). As the experimental signal included both electron and hole contributions, the electron component was isolated by extracting only the slow-relaxation component using a double-exponential fit, as in Fig. 2(c). The simulated valley polarization ratio was fitted by a single-exponential function to extract the relaxation time. Figure 4(f) shows the relaxation times obtained by the experiments and the Monte Carlo simulation as functions of the MIR fluence. The relaxation times are approximately 3 and 4 ps for the simulation and experiments, respectively, and are robust to the drive fluence in this range. Compared with the valley relaxation dynamics investigated in Fig. 2(d), the electron system driven by the MIR pulses corresponds well to the photoexcited one with the kinetic energy of 50-100 meV. While the valley relaxation in Fig. 2(d) was explained by the energy-dependent intervalley scattering rate alone, the present simulation further incorporates the field acceleration and intravalley scattering, and still reproduces the observed relaxation well.

Finally, we would like to underscore that the sub-picosecond complete valley reversal demonstrated in Fig. 4(d) promises novel ultrafast information processing, which may far exceed the switching technology in spintronics. Extensive efforts have been devoted to realizing ultrafast magnetization switching in the picosecond to sub-picosecond regime using optical pulses [43–45] and in the 10-ps regime using electrical pulses [46–50]. However, for information processing, a key figure of merit is not the switching speed itself but the maximum repetition rate of the operation. In spin systems, the repetition rate is severely limited by heat accumulation, which causes demagnetization or, in the worst case, device failure. Therefore, to increase the repetition rate, the energy required per bit written must be suppressed to reduce total energy consumption. Very recently, ultralow power switching of an antiferromagnet driven by 40-ps current pulses was demonstrated [50]; however, the suggested energy cost of ~1 fJ per bit relies on scaling down the device size to a 30 × 30 × 10 $nm^3$ dot. In stark contrast, the energy density imparted to the system per pulse for the valley switching in Ge presented here can be evaluated as $\int \sigma_{xx} E_{\mathrm{MIR}}^2 dt \sim 1$ aJ/$\mu m^3$ under the condition in Fig. 4(d) (see Supplemental Material [29]). Note that this estimation is based on the Drude model, which is not accurate for describing the intraband electron response in the MIR range. Nonetheless, ultralow energy consumption can still be expected for the valley switching. It highlights the direct coupling between the electric field and charge carriers via intraband acceleration without additional carrier excitation, in contrast to the difficulty in electrically controlling spins. While the valley switching was demonstrated here in the nearly transparent MIR region, similar valley switching is also anticipated to occur with a much weaker field using lower-frequency, picosecond (THz) electric pulses on chip because the acquired kinetic energy $U_p$ of electrons is proportional to $E_{\mathrm{MIR}}^2/\omega^2$.

In conclusion, time-resolved THz polarimetry, analyzed with quantitative theoretical calculations, revealed a valley lifetime of 10 ps in Ge at room temperature. Owing to this long lifetime, the valley polarization could be read out electrically as a DC transverse voltage. We also demonstrated sub-picosecond switching of valley polarization at an energy cost of only about 1 aJ/$\mu m^3$ per pulse. These results suggest that Ge is an ideal platform for valleytronic devices for ultrafast THz information processing. Because the generation, detection, and switching of valley polarization presented here rely only on the multiple anisotropic conduction-band valleys inherent to indirect-gap semiconductors, the present scheme should also be applicable to other materials. Future studies exploring strain engineering or low-dimensional structures could further enhance the valley polarization ratio and lifetime, ultimately opening new avenues toward practical room-temperature valleytronic logic and memory devices.

## Acknowledgements

A.M.S. thanks Yuto Fukushima for fruitful discussions. This work was supported by the JST FOREST (Grant Nos. JPMJFR2240 and JPMJFR213A), JSPS KAKENHI (Grant Nos. 24K00550 and 24K16988), JST PRESTO (Grant No. JPMJPR2006), and JST CREST (Grant No. JPMJCR20R4). R.M. acknowledges partial support from the MEXT Quantum Leap Flagship Program (MEXT Q-LEAP, Grant No. JPMXS0118068681). R.M. conceived the study. A.M.S. and Y.M. developed the pump–probe spectroscopy system with the help of T.F., K.A., J.Y. and R.M. C.K. and A.M.S. prepared the sample with the support of H.A. A.M.S. performed the pump–probe experiment and analyzed the data with the help of T.F. and Y.M. and also performed the transport experiment with the help of S.A. and T.I. A.M.S. conducted the Monte Carlo simulations with the help of S.A.S. All authors discussed the results. A.M.S. and R.M. wrote the manuscript with feedback from Y.M., T.I., and all coauthors.

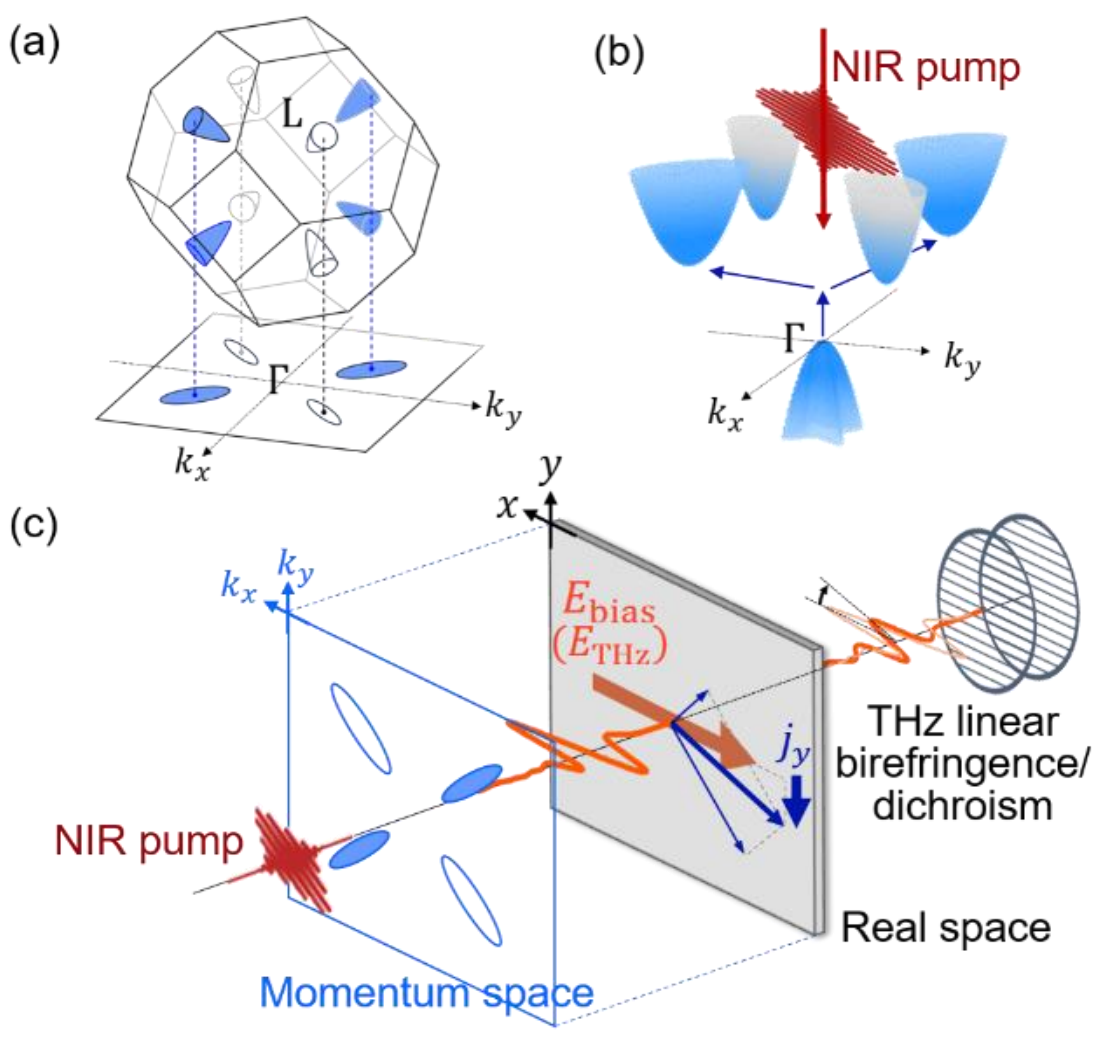

FIG. 1. (a) First Brillouin zone of Ge and its projection corresponding to the sample orientation in our experiment. (b) Schematic of light-induced valley polarization generated by the NIR light. (c) Schematic showing the generation of a photocurrent $j_y$ orthogonal to the bias field via the valley polarization and its detection as the THz linear birefringence and/or dichroism.

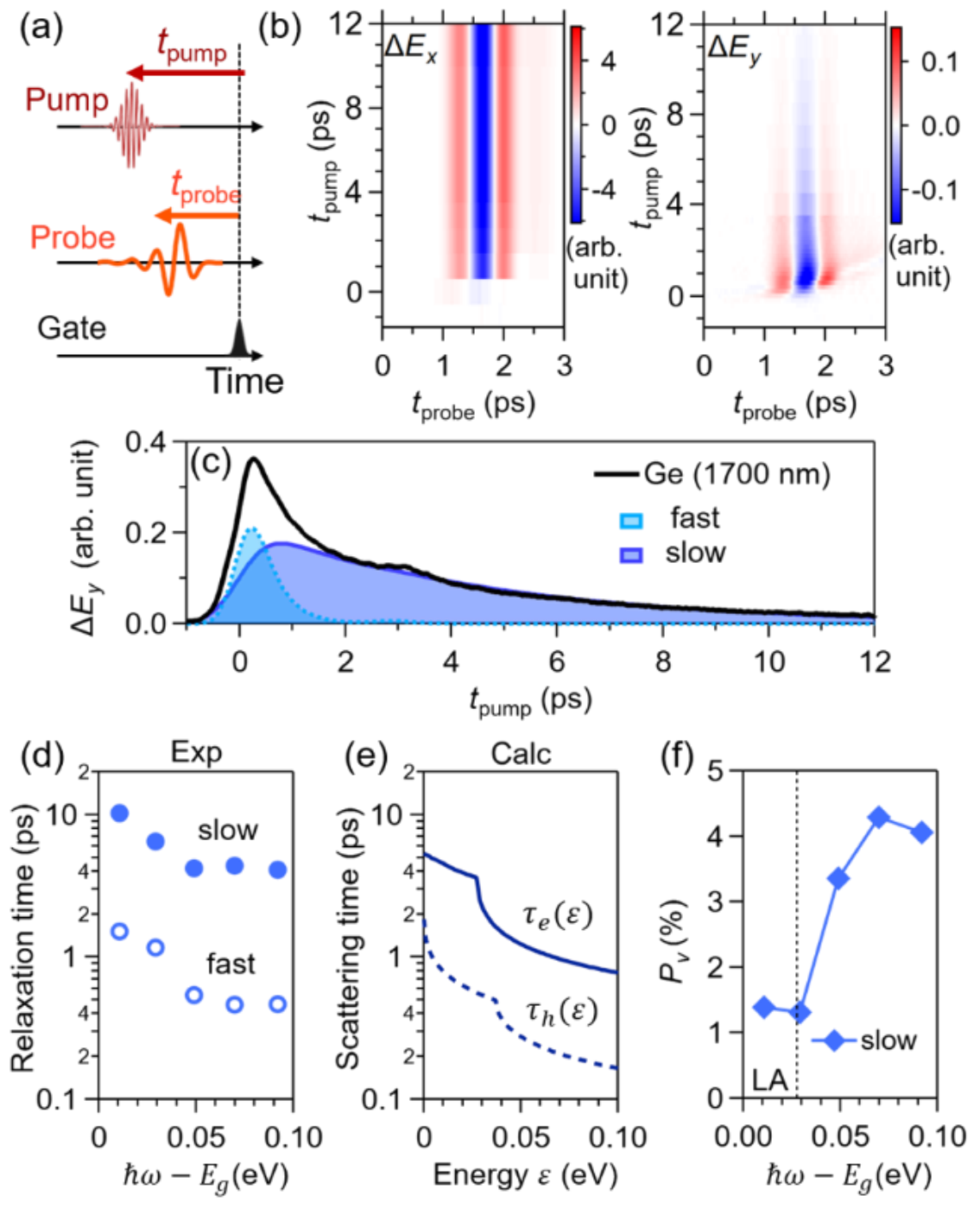

FIG. 2. (a) Time sequence of the NIR pump, THz probe, and gate pulses. (b) Two-dimensional plots of the THz field change parallel $[\Delta E_x(t_{\mathrm{probe}}, t_{\mathrm{pump}})]$ and perpendicular $[\Delta E_y(t_{\mathrm{probe}}, t_{\mathrm{pump}})]$ to the incident probe polarization. (c) $\Delta E_y(t_{\mathrm{pump}})$ at a fixed $t_{\mathrm{probe}}$ for a pump wavelength of 1700 nm (black line). Fast and slow relaxation components extracted from the double-exponential fitting (filled using light blue and dark blue colors). (d) Fast and slow relaxation times as functions of the excess photon energy $\hbar\omega - E_g$. (e) Solid and dashed lines represent the calculated electron intervalley scattering time $\tau_e$ and hole–phonon scattering time $\tau_h$, respectively. (f) Estimated valley polarization ratio $P_v$ as a function of the excess photon energy. The dashed line marks the LA phonon energy.

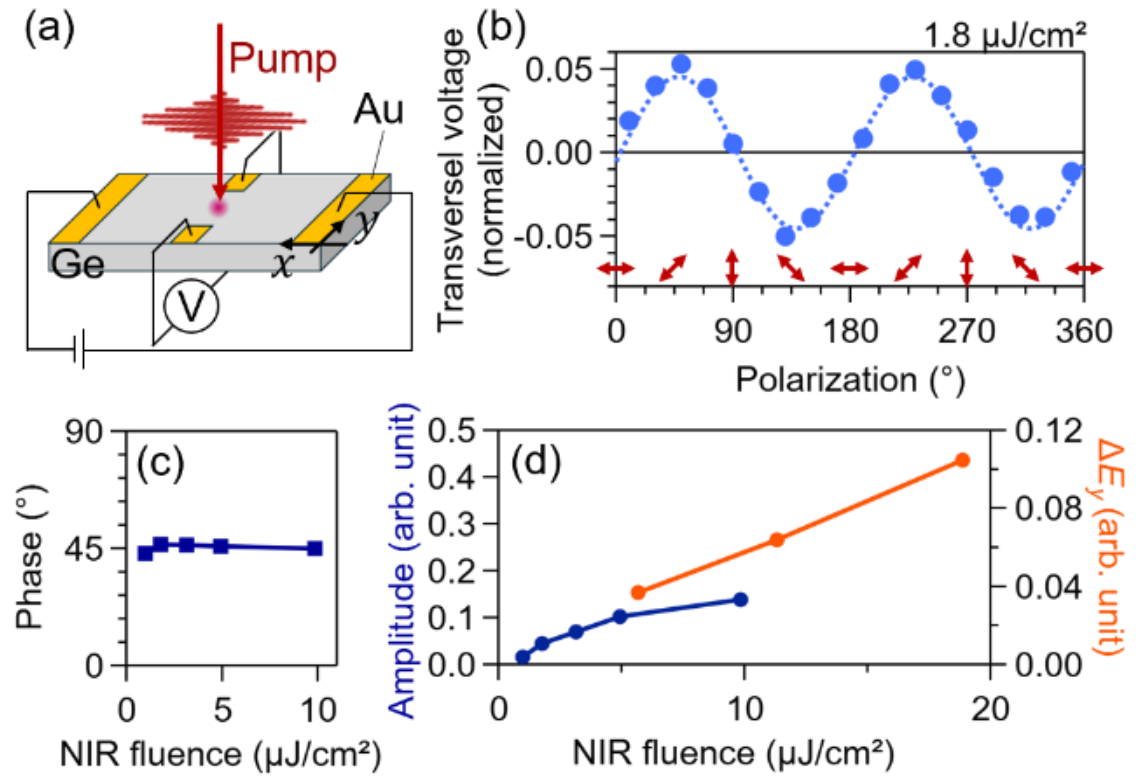

FIG. 3. (a) Schematic of the experimental setup for transport measurements. A bias voltage is applied along the positive $x$ axis. The transverse voltage is positive for currents flowing along the positive $y$ axis. (b) Transverse voltage depending on the pump polarization direction. Dashed curves denote a sine function fitting. Red arrows show the NIR polarization directions viewed along the light propagation path (from source to sample), where 0° aligns with the $x$ axis. (c) Phase corresponding to the oscillation peak of the transverse voltage as a function of the pump fluence. (d) Pump-fluence dependence of the oscillation amplitude of the transverse voltage (dark blue, left axis) and the THz polarization rotation $\Delta E_y$ (orange, right axis).

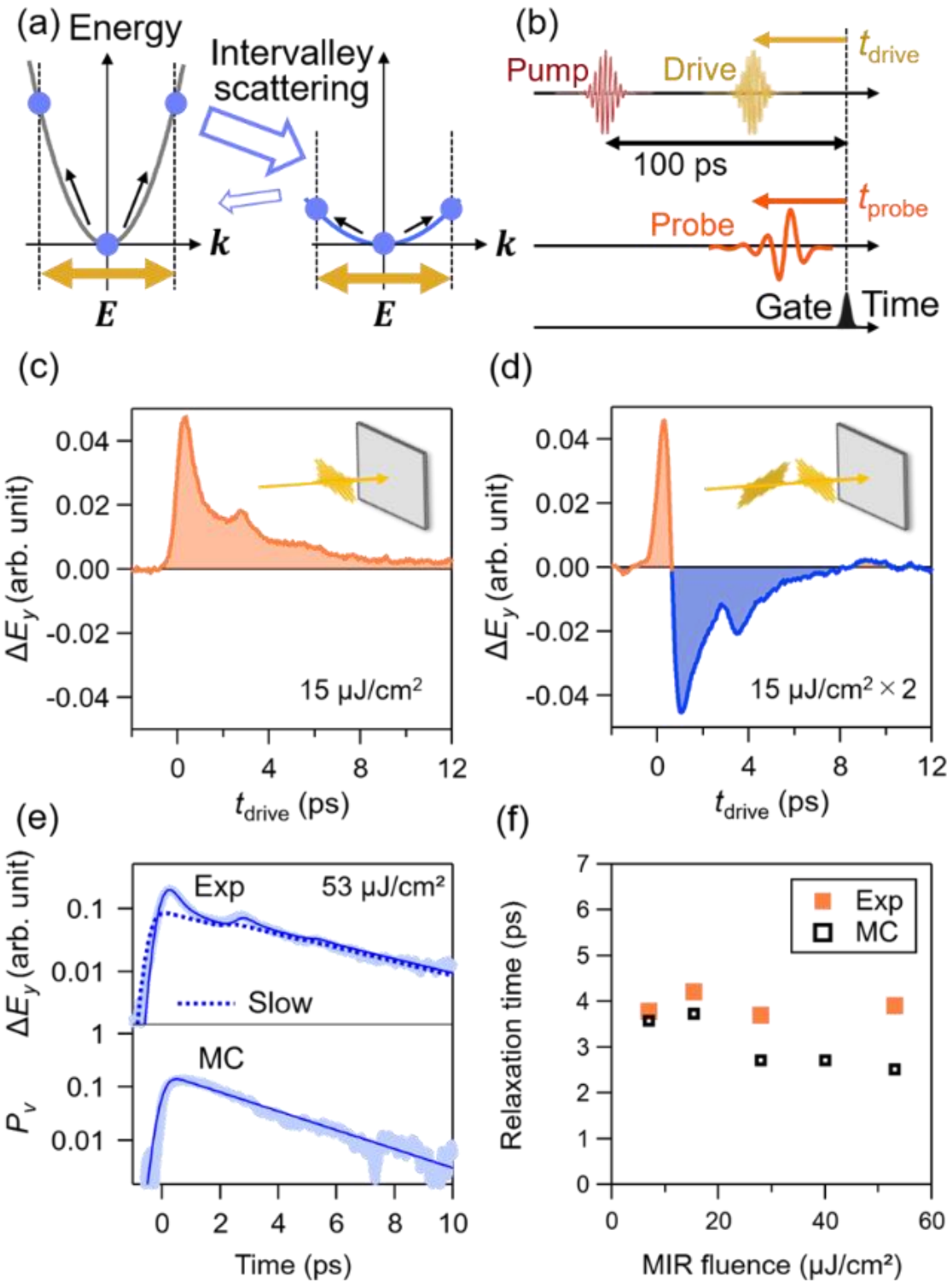

FIG. 4. (a) Schematic of carrier momentum shift induced by the off-resonant electric field (yellow arrow) within each valley, resulting in unbalanced intervalley scattering. (b) Pulse sequence of the NIR pump, MIR drive, THz probe, and gate pulses. $\Delta E_y$ as a function of $t_{\mathrm{drive}}$ for (c) single and (d) dual MIR pulses with mutually orthogonal polarizations. (e) Experimental THz polarization rotation dynamics together with the double-exponential fits (upper panel). The dashed curve indicates the slow decay component. Monte Carlo (MC) simulation of the valley polarization ratio $P_v$ with its single-exponential fit (lower panel). (f) Relaxation times obtained from the experiment (orange filled squares) and the MC simulation (black open) as functions of MIR fluence.